\documentclass[pdftex,twocolumn,epjc3]{svjour3}          % twocolumn

\smartqed  % flush right qed marks, e.g. at end of proof

\usepackage{flushend}
\usepackage[numbers,sort&compress]{natbib}
\usepackage[colorlinks,citecolor=blue,urlcolor=blue,linkcolor=blue]{hyperref}
\usepackage{graphicx}% Include figure files
\usepackage{subfig}
\usepackage{dcolumn}% Align table columns on decimal point
\usepackage{bm}% bold math
\usepackage{float}
\usepackage{xcolor}
\usepackage{amssymb}
\usepackage{amsmath}
\journalname{Eur. Phys. J. C}
\DeclareMathAlphabet{\mathcal}{OMS}{cmsy}{m}{n}
\DeclareSymbolFont{largesymbols}{OMX}{cmex}{m}{n}
\newcommand{\hz}{H({\mathrm z})}      % \rm: using roma as the form of a written or printed character

\newcommand{\densm}{\Omega_{\mathrm{m}}}

\newcommand{\hunit}{\mathrm{km \ s^{-1} \ Mpc^{-1}}}
\newcommand{\Om} {\ensuremath{\Omega_{\rm{m}}}}
\newcommand{\lya}{Lyman-$\alpha$}
\usepackage{array}
\makeatletter
\newcommand{\thickhline}{%
    \noalign {\ifnum 0=`}\fi \hrule height 1pt
    \futurelet \reserved@a \@xhline
}
\newcolumntype{"}{@{\hskip\tabcolsep\vrule width 1pt\hskip\tabcolsep}}
\makeatother
\begin{document}

\title{Cosmological model-independent test of $\Lambda$CDM with two-point diagnostic by the observational Hubble parameter data}

\author{Shu-Lei Cao\thanksref{addr1,addr2}
        \and
        Xiao-Wei Duan\thanksref{addr3} %etc.
        \and
        Xiao-Lei Meng\thanksref{addr2}
        \and
        Tong-Jie Zhang\thanksref{e1,addr1,addr2}
}

\thankstext{e1}{e-mail: tjzhang@bnu.edu.cn}

\institute{Dezhou University, Dezhou 253023, China\label{addr1}
          \and
          Department of Astronomy, Beijing Normal University, Beijing, 100875, China\label{addr2}
          \and
          Department of Astronomy, School of Physics, Peking University, Beijing 100871, China\label{addr3}
}

\date{Received: 08 February 2018 / Accepted: 09 April 2018 / Published online: 19 April 2018}

\maketitle

\begin{abstract}
Aiming at exploring the nature of dark energy (DE), we use forty-three observational Hubble parameter data (OHD) in the redshift range $0 < z \leqslant 2.36$ to make a cosmological model-independent test of the $\Lambda$CDM model with two-point $Omh^2(z_{2};z_{1})$ diagnostic. In $\Lambda$CDM model, with equation of state (EoS) $w=-1$, two-point diagnostic relation $Omh^2 \equiv \Om h^2$ is tenable, where $\Om$ is the present matter density parameter, and $h$ is the Hubble parameter divided by 100 $\hunit$. We utilize two methods: the weighted mean and median statistics to bin the OHD to increase the signal-to-noise ratio of the measurements. The binning methods turn out to be promising and considered to be robust. By applying the two-point diagnostic to the binned data, we find that although the best-fit values of $Omh^2$ fluctuate as the continuous redshift intervals change, on average, they are continuous with being constant within 1 $\sigma$ confidence interval. Therefore, we conclude that the $\Lambda$CDM model cannot be ruled out.
\end{abstract}

%\keywords{cosmological parameters --- Hubble constant --- Hubble parameter: error --- dark energy: error --- methods:statistical}

\section{Introduction}\label{sect:intro}

Over the past few decades, there have been a number of approaches proposed to quantitatively investigate the expansion history and structure
growth of the universe \citep[see][for recent reviews]{2008ARAA..46..385F,2013PhR...530...87W}. The observations of Type Ia supernovae (SNIa) \citep{1998AJ....116.1009R,1999ApJ...517..565P}
have provided ample evidence for an accelerating expansion to an increasing precision \citep{2012ApJ...746...85S}. Other complementary probes support that phenomenon,
including the baryon acoustic oscillation (BAO) measurements, the weak gravitational lensing, the abundance
of galaxy clusters \citep{2013ApJ...763..147B}, the cosmic microwave background (CMB) anisotropies, the linear growth of large-scale structure \citep{2002ApJ...572..140D}, and the Hubble constant $H_0$ \citep{2012ApJ...758...24F}. There are plenty of cosmological models raised to account for the acceleration phenomenon, yet the best-fit one is still uncertain.

The existence of dark energy (DE) with negative equation of state (EoS) parameter $w\equiv p_{\rm DE}/\rho_{\rm DE}$ is considered as a prevailing interpretation currently.
In this context, the most popular DE model being used remains to be the simple cosmological constant cold dark matter ($\Lambda \rm CDM$) model, with $w=-1$ at all times \citep{2011CoTPh..56..525L}.
However, the popularity of $\Lambda \rm CDM$ model does not cover the issue that it suffers from fine-tuning and coincidence problems \citep{1989RvMP...61....1W, 1999PhRvL..82..896Z}.
In addition, it is worth noticing about the possibility of DE with evolving EoS \citep{2017NatAs...1..627Z}, i.e. the dynamical DE models
($w=w(z)$), such as Quintessence \citep[$w>-1$,][]{1988PhRvD..37.3406R, 1988NuPhB.302..668W}, Phantom
\citep[$w<-1$,][]{2002PhLB..545...23C}, K-essence \citep[$w>-1~\textrm{or}~w<-1$,][]{2000PhRvL..85.4438A, 2001PhRvD..63j3510A},
and especially Quintom \citep[$w$ crossing -1,][]{2005PhLB..607...35F, 2006PhLB..634..101F} models. Nevertheless, it deserves more profound, physical explanations for all these models. Moreover, Zhao et al.
\cite{2017NatAs...1..627Z} find that an evolving DE can relieve the tensions presented among existing datasets within the $\Lambda$CDM framework.
Meanwhile, it is useful to introduce diagnostics based upon direct observations and capable of revealing dynamical features of DE. One of these diagnostics is
$Om(z)$, which is defined as a function of redshift $z$ \citep{2008PhRvD..78j3502S,2008PhRvL.101r1301Z}, i.e.,
\begin{equation}
   \label{eq:Om}
   Om(z) = \frac{\tilde{h}^2(z)-1}{(1+z)^3-1},
\end{equation}
with $\tilde{h}=\frac{H(z)}{H_{0}}$ and $H(z)$ denoting the Hubble expansion rate. $Om(z)$ has the property of being \Om{}
for $w=-1$ case. Moreover, Shafieloo et al. \cite{2012PhRvD..86j3527S} modified this diagnostic to accommodate two-point situations as follows,
\begin{equation}
   \label{eq:Om2points}
   Om(z_{2};z_{1}) = \frac{\tilde{h}^2(z_{2})-\tilde{h}^2(z_{1})}{(1+z_{2})^3-(1+z_{1})^3}.
\end{equation}
In this case, if $Om(z_{2};z_{1}) \equiv \Om$ held for any redshift intervals, it would substantiate the validity of $\Lambda$CDM. In other words, the measurements of $Om(z_{2};z_{1}) \ne \Om$ would imply a deviation from $\Lambda$CDM and the fact
that other DE models with an evolving EoS should be taken into account.

In this paper, we first introduce the measurements of the observational $H(z)$ data (OHD) and then exhibit the currently available data
sets in Section~\ref{sect:data}. In Section~\ref{sect:binOHD}, we apply two binning methods: the weighted mean and median statistics techniques to OHD, and obtain the binned OHD categorically.
In Section~\ref{sect:Omhapp}, based on the binned OHD, we test the $\Lambda$CDM model with two-point $Omh^2(z_2;z_1)$ diagnostic. Finally, we summarize our conclusions in Section~\ref{sect:conclu}.
%+++++++++++++++++++++

%============================= section 2 ===================================

\section{The observational $H(z)$ data sets}\label{sect:data}

The OHD can be used to constrain cosmological parameters because they are obtained from model-independent direct observations.
Until now, three methods have been developed to measure OHD: cosmic chronometers, radial BAO size methods \citep{2010AdAst2010E..81Z}, and gravitational waves \cite{2017arXiv170307923Y}.
Jimenez et al. \cite{2002ApJ...573...37J} first proposed that relative galaxy ages can be used to obtain $H(z)$
values and they reported one $H(z)$ measurement at $z \sim 0.1$ in their later work \citep{2003ApJ...593..622J}.
Simon et al. \cite{2005PhRvD..71l3001S} added additional eight $H(z)$ points in the redshift range between 0.17 and 1.75 from differential
ages of passively evolving galaxies, and further constrained the redshift dependence of the DE potential by reconstructing it as a
function of redshift. Later, Stern et al. \cite{2010JCAP...02..008S} provided two new determinations from red-envelope galaxies and then constrained
cosmological parameters including curvature through the joint analysis of CMB data. Furthermore,
Moresco et al. \cite{2012JCAP...08..006M} obtained eight new measurements of $H(z)$ from the differential spectroscopic evolution of early-type,
massive, red elliptical galaxies which can be used as standard cosmic chronometers. By applying the galaxy differential age method to SDSS DR7,
Zhang et al. \cite{2014RAA....14.1221Z} expanded the $H(z)$ data sample by four new points.
Taking advantage of near-infrared spectroscopy of high redshift galaxies, Moresco et al. \cite{2015MNRAS.450L..16M} obtained two measurements of $H(z)$.
Later, they gained five more latest $H(z)$ values \citep{Moresco2016}.
Rencently, Ratsimbazafy et al. \cite{2017MNRAS.467.3239R} provides one more measurement of $H(z)$ based on analysis of high quality spectra of Luminous Red Galaxies (LRGs) obtained with
the Southern African Large Telescope (SALT).

On the other side, $H(z)$ can also be extracted from the detection of radial BAO features. Gazta$\tilde{\rm n}$aga et al. \cite{2009MNRAS.399.1663G} first obtained two $H(z)$
data points using the BAO peak position as a standard ruler in the radial direction. Blake et al. \cite{2012MNRAS.425..405B} further combined the
measurements of BAO peaks and the Alcock-Paczynski distortion to find three other $H(z)$ results. Samushia et al. \cite{2013MNRAS.429.1514S} provided a
$H(z)$ point at $z = 0.57$ from the BOSS DR9 CMASS sample. Xu et al. \cite{2013MNRAS.431.2834X} used the BAO signals from the SDSS DR7 luminous red
galaxy sample to derive another observational $H(z)$ measurement. The $H(z)$ value determined based upon BAO features in
the \lya\ forest of SDSS-III quasars were presented by Delubac et al. \cite{2015AA...574A..59D} and Font-Ribera et al. \cite{2014JCAP...05..027F}, which are the farthest precisely observed $H(z)$ results so far. Alam et al. \cite{2017MNRAS.470.2617A} obtained three $H(z)$ measurements with cosmological analysis of the DR12 galaxy sample.

Moreover, Liu et al. \cite{2017arXiv170307923Y} present a new method of measuring Hubble parameter by using the anisotropy of luminosity distance, and the analysis of gravitational wave of neutron star binary system.

\begin{figure}
    \includegraphics[width=0.5\textwidth]{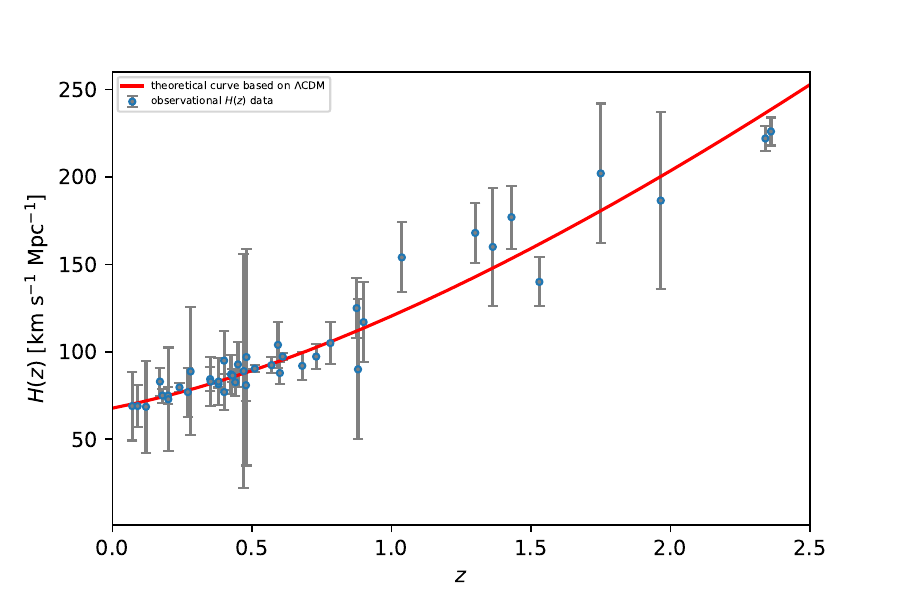}
    \caption{The full OHD set and theoretical curve of standard flat $\Lambda$CDM model. The OHD are represented by
    gray points along with the uncertainties. The red curve shows theoretical $H(z)$ evolution based upon the adopted
    $\Lambda$CDM model which is described in Section~\ref{sect:data}.}
    \label{fig:hzobs}
\end{figure}

After evaluating these data points from \cite{2017MNRAS.471L..82M,2017ApJ...835...26F}, we combine these 43 OHD and present them in Table~\ref{tab:Hzd} and mark them in Figure~\ref{fig:hzobs}. Note that it is obvious that the cosmic chronometer method is completely model-independent, and one may misjudge the radial BAO method to be dependent of model since they involve in some fiducial $\Lambda$CDM models. However, in fact, the fiducial models cannot affect the results as mentioned in the references (e.g., see P. 5 of Alam et al. \cite{2017MNRAS.470.2617A}). Moreover, the three $H(z)$ measurements taken from Blake et al. \cite{2012MNRAS.425..405B} are correlated with each other, and also, the three measurements of Alam et al. \cite{2017MNRAS.470.2617A} are correlated. This fact will affect the choice of binning range afterwards.

We use a $\Lambda$CDM model with no curvature term to compare theoretical values of Hubble parameter with the OHD results, with
the Hubble parameter given by
\begin{equation}
  H(z) = H_0 \sqrt{\Om (1+z)^3+1-\Om},
\end{equation}
where cosmological parameters take values from the Planck temperature power spectrum measurements \cite{2016AA...594A..13P}. The
best fit value of $H_0$ is 67.81 $\hunit$, and $\densm$ is 0.308. The theoretical computation of $H(z)$ based upon this
$\Lambda$CDM is also shown in Figure~\ref{fig:hzobs}.

Being independent observational data, $H(z)$ determinations have been frequently used in cosmological research. One of the leading
purposes is using them to constrain DE. Jimenez \& Loeb \cite{2002ApJ...573...37J} first proposed that $H(z)$ measurements can be used to
constrain DE EoS at high redshifts. Simon et al. \cite{2005PhRvD..71l3001S} derived constraints on DE potential using $H(z)$ results and
supernova data. Samushia \& Ratra \cite{2006ApJ...650L...5S} began applying these measurements to constraining cosmological parameters in various DE
models. In the meanwhile, DE evolution came into its own as an active research field in the last twenty years
\citep{2000IJMPD...9..373S,2001LRR.....4....1C,2003RvMP...75..559P,2006IJMPD..15.1753C,2012PhR...513....1C}. To sum up, the OHD
are proved to be very promising towards understanding the nature of DE.

\begin{table}[htp]
\centering
\setlength{\tabcolsep}{0.5mm}{
\begin{tabular}{lccc}
\hline
{$z$}   & $H(z)$ & Method & Ref.\\
\hline
$0.0708$   &  $69.0\pm19.68$      &  I    &  Zhang et al. (2014)-\cite{2014RAA....14.1221Z}   \\
      $0.09$       &  $69.0\pm12.0$        &  I    &  Jimenez et al. (2003)-\cite{2003ApJ...593..622J}   \\
      $0.12$       &  $68.6\pm26.2$        &  I    &  Zhang et al. (2014)-\cite{2014RAA....14.1221Z}   \\
      $0.17$       &  $83.0\pm8.0$          &  I    &  Simon et al. (2005)-\cite{2005PhRvD..71l3001S}     \\
      $0.179$     &  $75.0\pm4.0$          &  I    &  Moresco et al. (2012)-\cite{2012JCAP...08..006M}     \\
      $0.199$     &  $75.0\pm5.0$          &  I    &  Moresco et al. (2012)-\cite{2012JCAP...08..006M}     \\
      $0.2$         &  $72.9\pm29.6$        &  I    &  Zhang et al. (2014)-\cite{2014RAA....14.1221Z}   \\
      $0.240$     &  $79.69\pm2.65$      &  II   &  Gazta$\tilde{\rm{n}}$aga et al. (2009)-\cite{2009MNRAS.399.1663G}   \\
      $0.27$       &  $77.0\pm14.0$        &  I    &    Simon et al. (2005)-\cite{2005PhRvD..71l3001S}   \\
      $0.28$       &  $88.8\pm36.6$        &  I    &  Zhang et al. (2014)-\cite{2014RAA....14.1221Z}   \\
      $0.35$       &  $84.4\pm7.0$          &  II   &   Xu et al. (2013)-\cite{2013MNRAS.431.2834X}  \\
      $0.352$     &  $83.0\pm14.0$        &  I    &  Moresco et al. (2012)-\cite{2012JCAP...08..006M}   \\
      $0.38$      &  $81.5\pm1.9$        & II    &   Alam et al. (2016)-\cite{2017MNRAS.470.2617A}      \\
      $0.3802$     &  $83.0\pm13.5$        &  I    &  Moresco et al. (2016)-\cite{Moresco2016}   \\
      $0.4$         &  $95\pm17.0$           &  I    &  Simon et al. (2005)-\cite{2005PhRvD..71l3001S}     \\
      $0.4004$     &  $77.0\pm10.2$        &  I    &  Moresco et al. (2016)-\cite{Moresco2016}   \\
      $0.4247$     &  $87.1\pm11.2$        &  I    &  Moresco et al. (2016)-\cite{Moresco2016}   \\
      $0.43$     &  $86.45\pm3.68$        &  II   &  Gazta$\tilde{\rm{n}}$aga et al. (2009)-\cite{2009MNRAS.399.1663G}   \\
      $0.44$       & $82.6\pm7.8$           &  II   &  Blake et al. (2012)-\cite{2012MNRAS.425..405B}  \\
      $0.4497$     &  $92.8\pm12.9$        &  I    &  Moresco et al. (2016)-\cite{Moresco2016}   \\
      $0.47$      &   $89\pm67$           &   I    &  Ratsimbazafy et al. (2017)-\cite{2017MNRAS.467.3239R}    \\
      $0.4783$     &  $80.9\pm9.0$        &  I    &  Moresco et al. (2016)-\cite{Moresco2016}   \\
      $0.48$       &  $97.0\pm62.0$        &  I    &  Stern et al. (2010)-\cite{2010JCAP...02..008S}     \\
      $0.51$      &  $90.4\pm1.9$        & II    &   Alam et al. (2016)-\cite{2017MNRAS.470.2617A}      \\
      $0.57$       &  $92.4\pm4.5$          &  II   &  Samushia et al. (2013)-\cite{2013MNRAS.429.1514S}   \\
      $0.593$     &  $104.0\pm13.0$      &  I    &  Moresco et al. (2012)-\cite{2012JCAP...08..006M}   \\
      $0.6$         &  $87.9\pm6.1$          &  II   &  Blake et al. (2012)-\cite{2012MNRAS.425..405B}   \\
      $0.61$      &  $97.3\pm2.1$        & II    &   Alam et al. (2016)-\cite{2017MNRAS.470.2617A}      \\
      $0.68$       &  $92.0\pm8.0$          &  I    &  Moresco et al. (2012)-\cite{2012JCAP...08..006M}   \\
      $0.73$       &  $97.3\pm7.0$          &  II   &  Blake et al. (2012)-\cite{2012MNRAS.425..405B}  \\
      $0.781$     &  $105.0\pm12.0$      &  I    &  Moresco et al. (2012)-\cite{2012JCAP...08..006M}   \\
      $0.875$     &  $125.0\pm17.0$      &  I    &  Moresco et al. (2012)-\cite{2012JCAP...08..006M}   \\
      $0.88$       &  $90.0\pm40.0$        &  I    &  Stern et al. (2010)-\cite{2010JCAP...02..008S}     \\
      $0.9$         &  $117.0\pm23.0$      &  I    &  Simon et al. (2005)-\cite{2005PhRvD..71l3001S}  \\
      $1.037$     &  $154.0\pm20.0$      &  I    &  Moresco et al. (2012)-\cite{2012JCAP...08..006M}   \\
      $1.3$         &  $168.0\pm17.0$      &  I    &  Simon et al. (2005)-\cite{2005PhRvD..71l3001S}     \\
      $1.363$     &  $160.0\pm33.6$      &  I    &  Moresco (2015)-\cite{2015MNRAS.450L..16M}  \\
      $1.43$       &  $177.0\pm18.0$      &  I    &  Simon et al. (2005)-\cite{2005PhRvD..71l3001S}     \\
      $1.53$       &  $140.0\pm14.0$      &  I    &  Simon et al. (2005)-\cite{2005PhRvD..71l3001S}     \\
      $1.75$       &  $202.0\pm40.0$      &  I    &  Simon et al. (2005)-\cite{2005PhRvD..71l3001S}     \\
      $1.965$     &  $186.5\pm50.4$      &  I    &   Moresco (2015)-\cite{2015MNRAS.450L..16M}  \\
      $2.34$       &  $222.0\pm7.0$        &  II   &  Delubac et al. (2015)-\cite{2015AA...574A..59D}   \\
      $2.36$       &  $226.0\pm8.0$       &   II   &  Font-Ribera et al. (2014)-\cite{2014JCAP...05..027F}    \\
\hline
\end{tabular}}
\caption{\label{tab:Hzd} The current available OHD dataset. Where I and II represent the cosmic chronometer and the radial BAO size method, respectively, and $H(z)$ is in units of $\hunit$ here.}
\end{table}

In the next section, we will bin the OHD in Table~\ref{tab:Hzd} by using two binning techniques.
%++++++++++++++++++++++++++

%============================= section 3 ===================================

\section{Binning OHD}\label{sect:binOHD}

As stated by Farooq et al. \cite{2013PhLB..726...72F,2017ApJ...835...26F}, there are two techniques: the weighted mean and median statistics,
which can be used to bin Hubble parameter measurements. In \cite{2017ApJ...835...26F}, they listed two reasons to compute ``average'' $H(z)$ values for bins in redshift space.
On the one hand, the weighted mean technique can indicate whether the original data have error bars inconsistent with Gaussianity.
On the other hand, the binned data can more clearly visually illustrate tendencies as a function of redshift, without the assumption of a particular cosmological model.

As for the weighted mean technique, the ideal choice should be a trade-off between bin size and number of measurements per bin that maximizes both quantities. In order to avoid correlations at one bin, we choose 3-4, 4-5, 5-6, and 5-6-7 measurements per bin, which separates the correlated data, and the last four measurements are binned by twos for all the cases.

According to Podariu et al. \cite{2001ApJ...559....9P}, the weighted mean of $H(z)$ is given by

\begin{equation}\label{eq:Hzbar}
   \overline{H}(z) = \frac{\sum^N_{i=1} \left(H(z_i)/\sigma^2_i\right)}{\sum^N_{i=1} 1/\sigma^2_i},
\end{equation}
where $H(z_i)$ and $\sigma_i$ stand for the Hubble parameter data and the standard deviation for $i=1,2,...,N$ measurements in the binning redshift range. Similarly, the corresponding
weighted bin redshift $\overline{z}$ and weighted error $\overline{\sigma}$ are as follows,

\begin{equation}\label{eq:zbar}
  \overline{z} = \frac{\sum^N_{i=1} \left(z_i/\sigma^2_i\right)}{\sum^N_{i=1} 1/\sigma^2_i},
\end{equation}
and
\begin{equation}\label{eq:ws}
  \overline{\sigma} = \sqrt{\frac{1}{\sum^N_{i=1} 1/\sigma^2_i}}.
\end{equation}
The goodness of fit for each bin, the reduced $\chi^2_{\nu}$, can be expressed as

\begin{equation}\label{eq:chi}
  \chi^2_{\nu}=\frac{1}{N-1}\sum^N_{i=1}\frac{\left(H(z_i)-\overline{H}(z)\right)^2}{\sigma^2_i},
\end{equation}
where the expected value and error of $\chi_{\nu}$ are unity and \\
$1/\sqrt{2(N-1)}$. Thus the number of standard deviations which $\chi_{\nu}$ deviates from unity for each bin is presented as

\begin{equation}\label{eq:chi}
  N_{\sigma}=|\chi_{\nu}-1|\sqrt{2(N-1)}.
\end{equation}
Non-Gaussian measurements, the presence of unaccounted for systematic errors, or correlations between measurements can result in large $N_{\sigma}$.
The weighted mean results for the binned $H(z)$ measurements are listed in Table \ref{tab:wmbin}, where the $N_{\sigma}$ values are considerably small for all bins, just like
results of Farooq et al. \cite{2017ApJ...835...26F}, hence indicating that the 43 OHD are not inconsistent with Gaussianity.

\begin{table}[htp]
\centering
\caption{\label{tab:wmbin} Weighted Mean Results for 43 Redshift Measurements, where the unit of $\hz$ is $\hunit$.}
\setlength{\tabcolsep}{0.7mm}{
\begin{tabular}{lcccccc}
\thickhline
\hline
{Bin} & N  &  $z$  & $H(z)$ & $H(z)$\ (1$\sigma$ Range) & $H(z)$\ (2$\sigma$ Range) & $N_{\sigma}$\\
\thickhline
\multicolumn{7}{c}{3 or 4 Measurements per Bin}  \\
\thickhline
1 & 3 & 0.0895 & 68.9 & 59.4-78.5 & 49.9-88.0 & 1.98 \\
2 & 4 & 0.1847 & 76.0 & 73.1-78.9 & 70.2-81.8 & 1.12 \\
3 & 3 & 0.2412 & 79.6 & 77.0-82.2 & 74.4-84.8 & 1.56 \\
4 & 4 & 0.3776 & 81.7 & 79.9-83.5 & 78.1-85.3 & 1.85 \\
5 & 3 & 0.4095  & 83.8 & 76.9-90.7 & 70.0-97.6 & 0.61 \\
6 & 3 & 0.4329  & 86.2 & 83.0-89.4 & 79.8-92.6 & 1.02 \\
7 & 4 & 0.5086  & 90.0 & 88.1-91.9 & 86.2-93.8 & 0.98 \\
8 & 4 & 0.6024  & 95.8 & 94.0-97.6 & 92.2-99.4 & 0.06 \\
9 & 3 & 0.7201 & 96.6 & 91.8-101.4 & 87.0-106.2 & 0.71 \\
10 & 4 & 0.9287  & 129.2 & 118.3-140.1 & 107.4-151.0 & 0.07 \\
11 & 4 & 1.4301 & 158.3 & 149.4-167.2 & 140.5-176.1 & 0.05 \\
12 & 2 & 1.8331  & 196.0 & 164.7-227.3 & 133.4-258.6 & 1.07 \\
13 & 2 & 2.3487  & 223.7 & 218.4-229.0 & 213.1-234.3 & 0.88 \\
\thickhline
\multicolumn{7}{c}{4 or 5 Measurements per Bin}  \\
\thickhline
1 & 5 & 0.1664 & 75.7 & 72.4-79.0 & 69.1-82.3 & 1.18 \\
2 & 5 & 0.2321 & 78.6 & 76.3-80.9 & 74.0-83.2 & 1.55 \\
3 & 4 & 0.3776 & 81.7 & 79.9-83.5 & 78.1-85.3 & 1.85 \\
4 & 5 & 0.4276 & 85.4 & 82.4-88.4 & 79.4-91.4 & 1.26 \\
5 & 5 & 0.5074 & 90.1 & 88.3-91.9 & 86.5-93.7 & 1.33 \\
6 & 5 & 0.6061  & 95.6 & 93.8-97.4 & 92.0-99.2 & 0.23 \\
7 & 5 & 0.7681  & 102.8 & 97.3-108.3 & 91.8-113.8 & 0.45 \\
8 & 5 & 1.3647  & 157.5 & 149.3-165.7 & 141.1-173.9 & 0.32 \\
9 & 2 & 1.8331  & 196.0 & 164.7-227.3 & 133.4-258.6 & 1.07 \\
10 & 2 & 2.3487  & 223.7 & 218.4-229.0 & 213.1-234.3 & 0.88 \\
\thickhline
\multicolumn{7}{c}{5 or 6 Measurements per Bin}  \\
\thickhline
1 & 5 & 0.1664 & 75.7 & 72.4-79.0 & 69.1-82.3 & 1.18 \\
2 & 5 & 0.2321 & 78.6 & 76.3-80.9 & 74.0-83.2 & 1.55 \\
3 & 6 & 0.3785 & 81.7 & 80.0-83.5 & 78.2-85.3 & 1.75 \\
4 & 6 & 0.4276 & 85.7 & 82.8-88.6 & 79.9-91.5 & 1.89 \\
5 & 5 & 0.5263 & 90.7 & 89.0-92.4 & 87.3-94.1 & 1.13 \\
6 & 6 & 0.6312  & 97.5 & 95.6-99.4 & 93.7-101.3 & 0.51 \\
7 & 6 & 1.3128  & 153.0 & 145.3-160.7 & 137.6-168.4 & 0.28 \\
8 & 2 & 1.8331  & 196.0 & 164.7-227.3 & 133.4-258.6 & 1.07 \\
9 & 2 & 2.3487  & 223.7 & 218.4-229.0 & 213.1-234.3 & 0.88 \\
\thickhline
\multicolumn{7}{c}{5, 6 or 7 Measurements per Bin}  \\
\thickhline
1 & 7 & 0.1767 & 75.4 & 72.6-78.2 & 69.8-81.0 & 1.80 \\
2 & 7 & 0.3333 & 81.1 & 79.6-82.6 & 78.1-84.1 & 2.27 \\
3 & 7 & 0.4288 & 85.8 & 82.9-88.7 & 80.0-91.6 & 1.71 \\
4 & 6 & 0.5247 & 90.4 & 88.8-92.0 & 87.1-93.7 & 0.89 \\
5 & 7 & 0.6331 & 97.7 & 95.8-99.6 & 93.9-101.5 & 0.55 \\
6 & 5 & 1.3647  & 157.5 & 149.3-165.7 & 141.1-173.9 & 0.32 \\
7 & 2 & 1.8331  & 196.0 & 164.7-227.3 & 133.4-258.6 & 1.07 \\
8 & 2 & 2.3487  & 223.7 & 218.4-229.0 & 213.1-234.3 & 0.88 \\
\hline
\end{tabular}}
\end{table}

Since the median statistics technique originally proposed by Gott et al.\cite{2001ApJ...549....1G} has a prerequisite which assumes that measurements of a given quantity are independent
and that there are no systematic effects, and as previously mentioned the correlative measurements of OHD from Blake et al. \cite{2012MNRAS.425..405B} and Alam et al. \cite{2017MNRAS.470.2617A} may contaminate the results,
we decide to remove these data for the sake of purity. While other measurements are all uncorrelated and independent with each other, they are more sustainable for evaluation and free of negative influence on the binning results. Table \ref{tab:msbin} displays the results from median statistics technique. After assuming that there is no overall systematic error in the reduced OHD as a whole and all the remaining measurements are independent, it is convenient to use the median statistics to combine the OHD. As the number of the measurements increases and approaches to infinity, the median can be presented as a true value, therefore, this technique has the merits of reducing the effect of outliers of a set of measurements on the estimate of a true median value. Nevertheless, although OHD are a bit of short in quantity as opposed to the large amount of measurements needed to reveal the true value, we still employ this technique for comparison purpose. If $N$ measurements $M_i$ (where $i = 1,2,...,N$) are considered, the probability of finding the true median between values $M_i$ and $M_{i+1}$ is \cite{2001ApJ...549....1G}

\begin{equation}\label{eq:prob}
  P_i=\frac{2^{-N}N!}{i!(N-i)!},
\end{equation}
where $N!$ represents the factorial of $N$. After applying this technique to the reduced 37 OHD for binning, we obtain the binned results as listed in Table \ref{tab:msbin}, applying the same binning scheme presented above. The results seem reasonable, but the precision is less than the weighted mean results, which may be caused by the smaller amount of OHD.

\begin{table}[htbp]
\centering
\caption{\label{tab:msbin} Median Statistics Results for 37 reduced Redshift Measurements, where the unit of $\hz$ is $\hunit$.}
\setlength{\tabcolsep}{1mm}{
\begin{tabular}{lccccc}
\thickhline
\hline
{Bin} & N  &  $z$  & $H(z)$ & $H(z)$\ (1$\sigma$ Range) & $H(z)$\ (2$\sigma$ Range)\\
\thickhline
\multicolumn{6}{c}{3 or 4 Measurements per Bin}  \\
\thickhline
1 & 3 & 0.09 & 69.0 & 49.3-88.7 & 29.6-108.4  \\
2 & 4 & 0.189 & 75.0 & 68.5-81.5 & 62.0-88.0  \\
3 & 3 & 0.27 & 79.7 & 65.7-93.7 & 51.7-107.7  \\
4 & 3 & 0.352 & 83.0 & 69.5-96.5 & 56-110.0  \\
5 & 4 & 0.4125  & 86.8 & 76.1-97.5 & 65.4-108.2  \\
6 & 4 & 0.474  & 90.9 & 53.5-128.4 & 16.0-165.8  \\
7 & 4 & 0.6365  & 98.2 & 88.2-108.2 & 78.2-118.2  \\
8 & 4 & 0.89  & 121.0 & 99.5-142.5 & 78.0-164.0  \\
9 & 4 & 1.3965 & 164.0 & 146.5-181.5 & 129.0-199.0  \\
10 & 4 & 2.1525  & 212.0 & 188.0-236.0 & 164.0-260.0  \\
\thickhline
\multicolumn{6}{c}{4 or 5 Measurements per Bin}  \\
\thickhline
1 & 5 & 0.12 & 69.0 & 57.0-81.0 & 45.0-93.0  \\
2 & 5 & 0.24 & 77.0 & 63.0-91.0 & 49.0-105.0  \\
3 & 5 & 0.3802 & 83.0 & 69.5-96.5 & 56.0-110.0  \\
4 & 5 & 0.4497 & 87.1 & 75.9-98.3 & 64.7-109.5  \\
5 & 5 & 0.593  & 97.0 & 85.0-109.0 & 73.0-121.0  \\
6 & 4 & 0.89  & 121.0 & 99.5-142.5 & 78.0-164.0  \\
7 & 4 & 1.3965 & 164.0 & 146.5-181.5 & 129.0-199.0  \\
8 & 4 & 2.1525  & 212.0 & 188.0-236.0 & 164.0-260.0  \\
\thickhline
\multicolumn{6}{c}{4, 5 or 6 Measurements per Bin}  \\
\thickhline
1 & 5 & 0.12 & 69.0 & 57.0-81.0 & 45.0-93.0  \\
2 & 5 & 0.24 & 77.0 & 63.0-91.0 & 49.0-105.0  \\
3 & 5 & 0.3802 & 83.0 & 69.5-96.5 & 56.0-110.0  \\
4 & 6 & 0.4598 & 88.1 & 76.0-100.2 & 63.9-112.3  \\
5 & 6 & 0.7305  & 98.2 & 85.7-110.7 & 73.2-123.2  \\
6 & 6 & 1.3315  & 157.0 & 138.0-176.0 & 119.0-195.0  \\
7 & 4 & 2.1525  & 212.0 & 188.0-236.0 & 164.0-260.0  \\
\thickhline
\multicolumn{6}{c}{4, 6 or 7 Measurements per Bin}  \\
\thickhline
1 & 7 & 0.17 & 72.9 & 60.9-84.9 & 48.9-96.9 \\
2 & 6 & 0.315 & 83.0 & 69.3-96.8 & 55.5-110.5 \\
3 & 7 & 0.43 & 87.1 & 75.9-98.3 & 64.7-109.5 \\
4 & 7 & 0.68 & 97.0 & 84.0-110.0 & 71.0-123.0 \\
5 & 6 & 1.3315  & 157.0 & 138.0-176.0 & 119.0-195.0  \\
6 & 4 & 2.1525  & 212.0 & 188.0-236.0 & 164.0-260.0  \\
\hline
\end{tabular}}
\end{table}

Even though the OHD obtained from the BAO method are model-independent, one can still be confused by the employed fiducial models. Therefore, to avoid the confusion and also for comparison purpose,
we decide to exclude the OHD from the BAO method and only consider the cosmic chronometer case, which also have the merit of being independent with each other. Table
\ref{tab:CCwmbin} and \ref{tab:CCmsbin} present the results from both binning methods. The results are all reasonable as well, and compared to the full binned OHD, the discrepancies are considerably small, which can be seen as evidence of the validity of the OHD derived from the BAO method. Since the different measurements per bin do not significantly affect the
results, we can acknowledge the robustness of the binning methods.

\begin{table}[htbp]
\centering
\caption{\label{tab:CCwmbin} Weighted Mean Results for Cosmic Chronometer Measurements, where the unit of $\hz$ is $\hunit$.}
\setlength{\tabcolsep}{0.6mm}{
\begin{tabular}{lcccccc}
\thickhline
\hline
{Bin} & N  &  $z$  & $H(z)$ & $H(z)$\ (1$\sigma$ Range) & $H(z)$\ (2$\sigma$ Range) & $N_{\sigma}$\\
\thickhline
\multicolumn{7}{c}{3 or 4 Measurements per Bin}  \\
\thickhline
1 & 3 & 0.0895 & 68.9 & 59.4-78.5 & 49.9-88.0 & 1.98 \\
2 & 4 & 0.1847 & 76.0 & 73.1-78.9 & 70.2-81.8 & 1.12 \\
3 & 3 & 0.3089 & 80.6 & 71.0-91.2 & 61.5-99.7 & 1.46 \\
4 & 4 & 0.4035 & 83.6 & 77.5-89.7 & 71.3-95.9 & 1.06 \\
5 & 4 & 0.4691  & 85.0 & 77.7-92.3 & 70.4-99.6 & 1.34 \\
6 & 3 & 0.6866  & 97.7 & 91.7-103.6 & 85.8-109.5 & 0.51 \\
7 & 3 & 0.8834  & 118.8 & 105.9-131.7 & 93.0-144.6 & 0.85 \\
8 & 4 & 1.2798  & 166.6 & 156.6-176.6 & 146.6-186.6 & 1.20 \\
9 & 3 & 1.58 & 149.3 & 136.5-162.1 & 123.7-174.9 & 0.33 \\
\thickhline
\multicolumn{7}{c}{3, 4 or 5 Measurements per Bin}  \\
\thickhline
1 & 5 & 0.1664 & 75.7 & 72.4-79.0 & 69.1-82.3 & 1.18 \\
2 & 5 & 0.2221 & 76.1 & 71.7-80.5 & 67.3-84.9 & 1.90 \\
3 & 5 & 0.412 & 85.3 & 79.8-90.8 & 74.2-96.4 & 1.17 \\
4 & 5 & 0.5897 & 90.1 & 84.7-85.5 & 79.3-100.9 & 0.70 \\
5 & 4 & 0.5074 & 111.4 & 102.6-120.2 & 93.8-129 & 0.86 \\
8 & 4 & 1.2798  & 166.6 & 156.6-176.6 & 146.6-186.6 & 1.20 \\
9 & 3 & 1.58 & 149.3 & 136.5-162.1 & 123.7-174.9 & 0.33 \\
\thickhline
\multicolumn{7}{c}{5 or 6 Measurements per Bin}  \\
\thickhline
1 & 5 & 0.1664 & 75.7 & 72.4-79.0 & 69.1-82.3 & 1.18 \\
2 & 5 & 0.2221 & 76.1 & 71.7-80.5 & 67.3-84.9 & 1.90 \\
3 & 5 & 0.412 & 85.3 & 79.8-90.8 & 74.2-96.4 & 1.17 \\
4 & 5 & 0.5897 & 90.1 & 84.7-85.5 & 79.3-100.9 & 0.70 \\
5 & 5 & 0.8622 & 118.3 & 110.2-126.4 & 102.2-134.4 & 0.36 \\
6 & 6 & 0.6312  & 161.1 & 152.5-169.7 & 143.9-178.3 & 0.16 \\
\hline
\end{tabular}}
\end{table}

\begin{table}[htbp]
\centering
\caption{\label{tab:CCmsbin} Median Statistics Results for Cosmic Chronometer Measurements, where the unit of $\hz$ is $\hunit$.}
\setlength{\tabcolsep}{1mm}{
\begin{tabular}{lccccc}
\thickhline
\hline
{Bin} & N  &  $z$  & $H(z)$ & $H(z)$\ (1$\sigma$ Range) & $H(z)$\ (2$\sigma$ Range)\\
\thickhline
\multicolumn{6}{c}{3 or 4 Measurements per Bin}  \\
\thickhline
1 & 3 & 0.09 & 69.0 & 49.3-88.7 & 29.6-108.4  \\
2 & 4 & 0.189 & 75.0 & 68.5-81.5 & 62.0-88.0  \\
3 & 3 & 0.28 & 83.0 & 69.0-97.0 & 55.0-111.0  \\
4 & 4 & 0.4002 & 85.1 & 72.7-97.4 & 60.4-109.8  \\
5 & 4 & 0.4741  & 90.9 & 53.5-128.4 & 16.0-165.8  \\
6 & 3 & 0.68  & 104.0 & 92.0-116.0 & 80.0-128.0  \\
7 & 3 & 0.88  & 117.0 & 94.0-140.0 & 71.0-163.0  \\
8 & 4 & 1.3315  & 164.0 & 145.0-183.0 & 126.0-202.0  \\
9 & 3 & 1.75 & 186.5 & 146.5-226.5 & 106.5-266.5  \\
\thickhline
\multicolumn{6}{c}{3, 4 or 5 Measurements per Bin}  \\
\thickhline
1 & 5 & 0.12 & 69.0 & 57.0-81.0 & 45.0-93.0  \\
2 & 5 & 0.27 & 77.0 & 63.0-91.0 & 49.0-105.0  \\
3 & 5 & 0.4004 & 87.1 & 74.2-100.0 & 61.3-112.9  \\
4 & 5 & 0.48 & 92.0 & 79.0-105.0 & 66.0-118.0  \\
5 & 4 & 0.8775  & 111.0 & 91.0-131.0 & 71.0-151.0  \\
6 & 4 & 1.3315  & 164.0 & 145.0-183.0 & 126.0-202.0  \\
7 & 3 & 1.75 & 172.5 & 146.5-226.5 & 106.5-266.5  \\
\thickhline
\multicolumn{6}{c}{5 or 6 Measurements per Bin}  \\
\thickhline
1 & 5 & 0.12 & 69.0 & 57.0-81.0 & 45.0-93.0  \\
2 & 5 & 0.27 & 77.0 & 63.0-91.0 & 49.0-105.0  \\
3 & 5 & 0.4004 & 87.1 & 74.2-100.0 & 61.3-112.9  \\
4 & 5 & 0.48 & 92.0 & 79.0-105.0 & 66.0-118.0  \\
5 & 5 & 0.88  & 117.0 & 97.0-137.0 & 77.0-157.0  \\
6 & 6 & 1.48  & 172.5 & 146.7-198.3 & 120.9-224.1  \\
\hline
\end{tabular}}
\end{table}

\section{Testing the $\Lambda$CDM model with $Omh^2(z_{2};z_{1})$ diagnostic}\label{sect:Omhapp}

In our analysis, the validity of $Omh^2(z_{2};z_{1})$ diagnostic can be tested using $H(z)$ results from cosmological independent measurements.
On the basis of the above section, we apply binned OHD from both the weighted mean technique and the median statistics technique to the two-point $Omh^2(z_{2};z_{1})$ diagnostic.

\begin{figure*}[htbp]
\centering
  \subfloat[]{%
    \includegraphics[width=6.5in,height=1.9in]{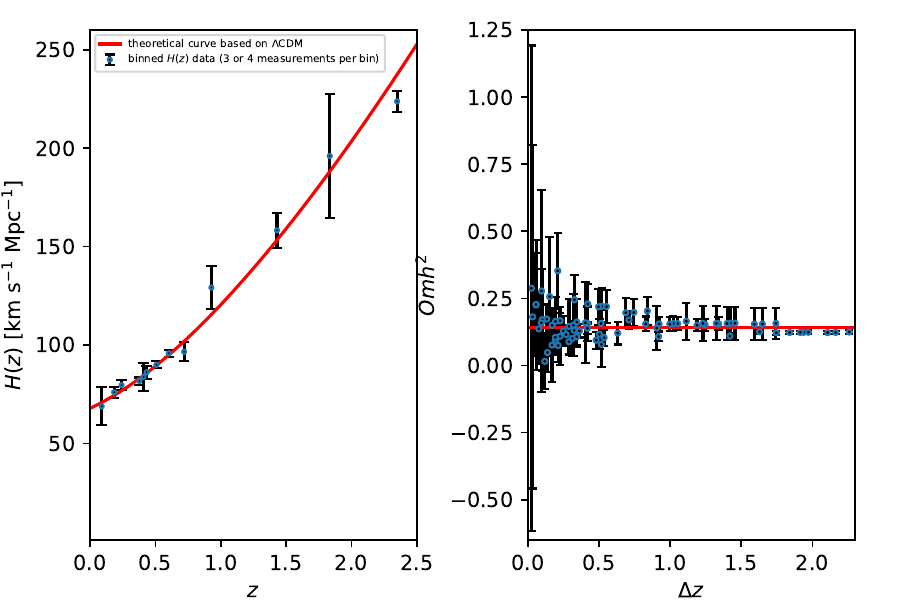}}\\
  \subfloat[]{%
    \includegraphics[width=6.5in,height=1.9in]{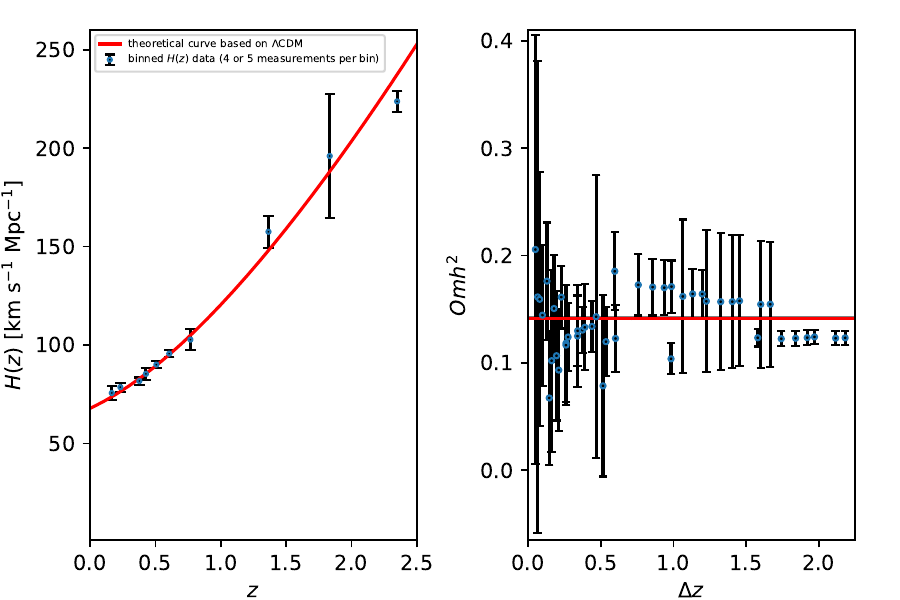}}\\
  \subfloat[]{%
    \includegraphics[width=6.5in,height=1.9in]{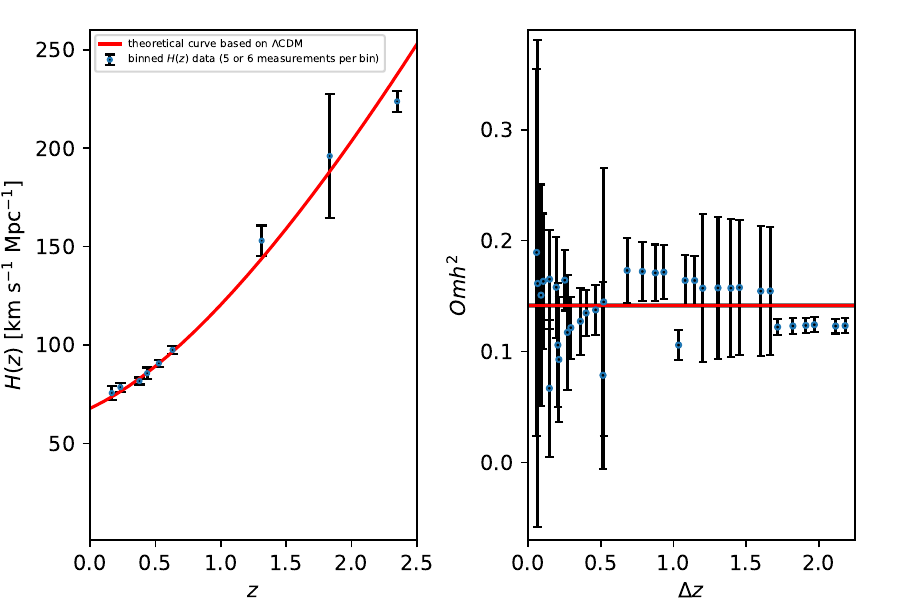}}\\
  \subfloat[]{%
    \includegraphics[width=6.5in,height=1.9in]{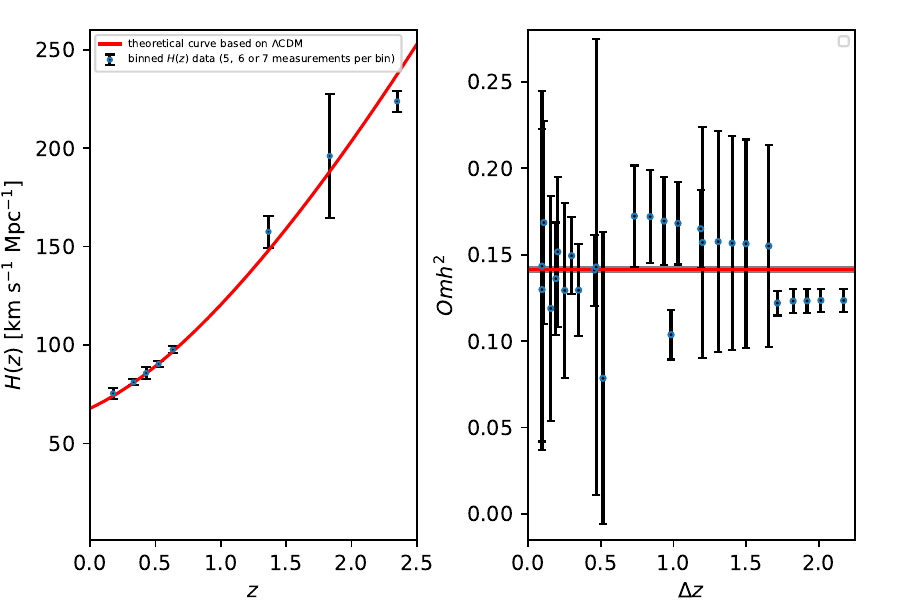}}\\
  \caption{Binned $H(z)$ data based on full OHD and its corresponding $Omh^2$ diagnostic values for the weighted mean technique. The panels (a), (b) (c) and (d) present the 3-4, 4-5, 5-6 and 5-6-7 measurements per bin case, respectively, where the red lines and grey shaded regions in the right panels represent the best-fit value of $Omh^2$ and its uncertainties retrieved from the Planck CMB data.}\label{fig:HzOmh1}
\end{figure*}

\begin{figure*}[htbp]
\centering
  \subfloat[]{%
    \includegraphics[width=6.5in,height=1.9in]{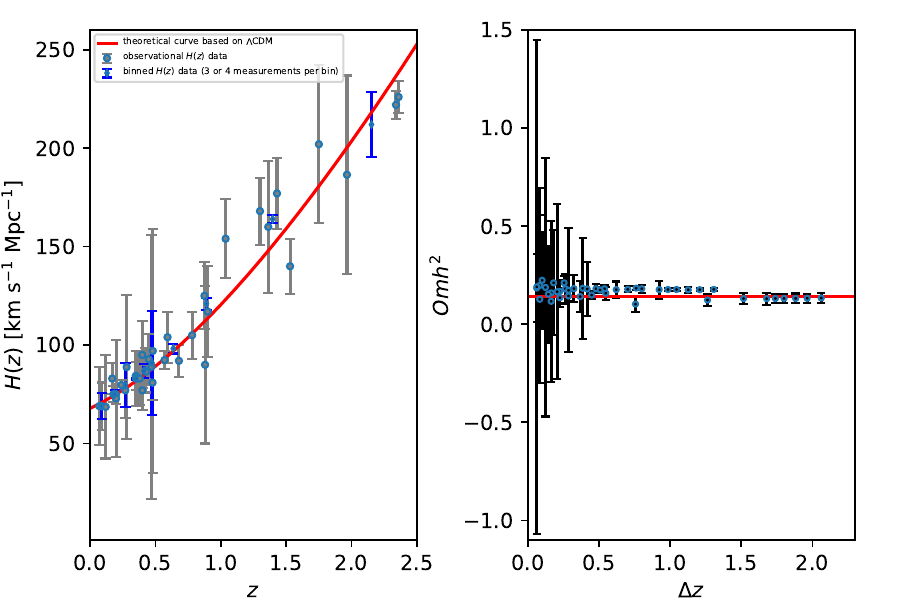}}\\
  \subfloat[]{%
    \includegraphics[width=6.5in,height=1.9in]{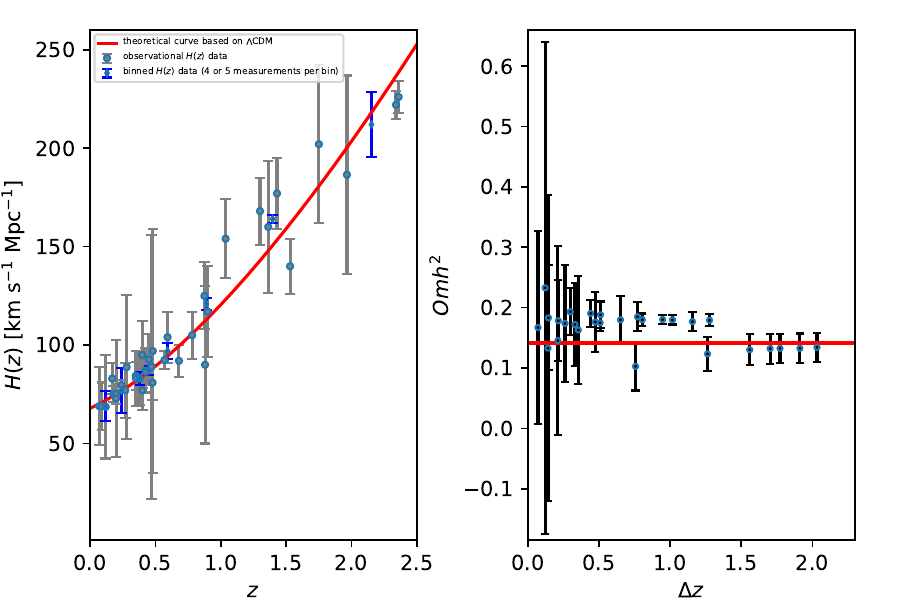}}\\
  \subfloat[]{%
    \includegraphics[width=6.5in,height=1.9in]{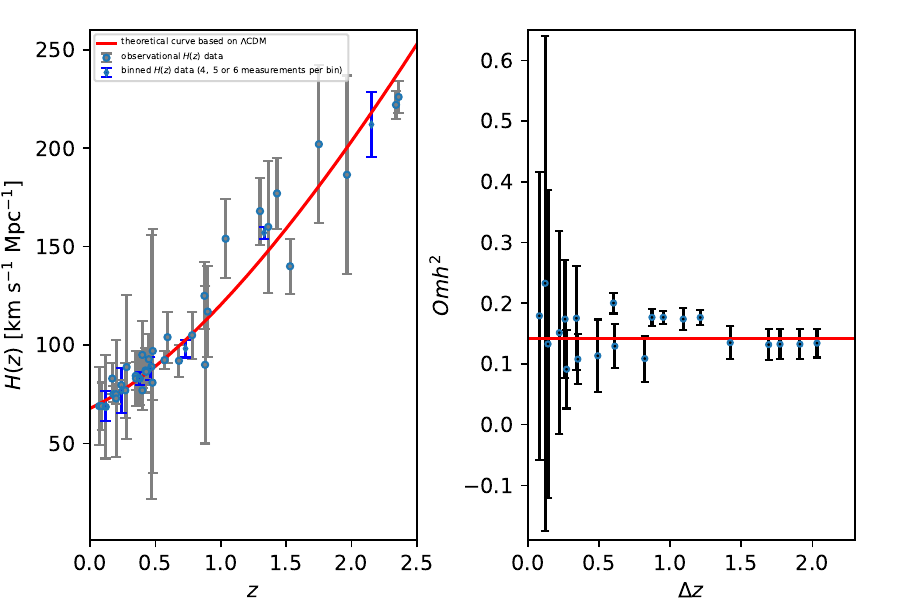}}\\
  \subfloat[]{%
    \includegraphics[width=6.5in,height=1.9in]{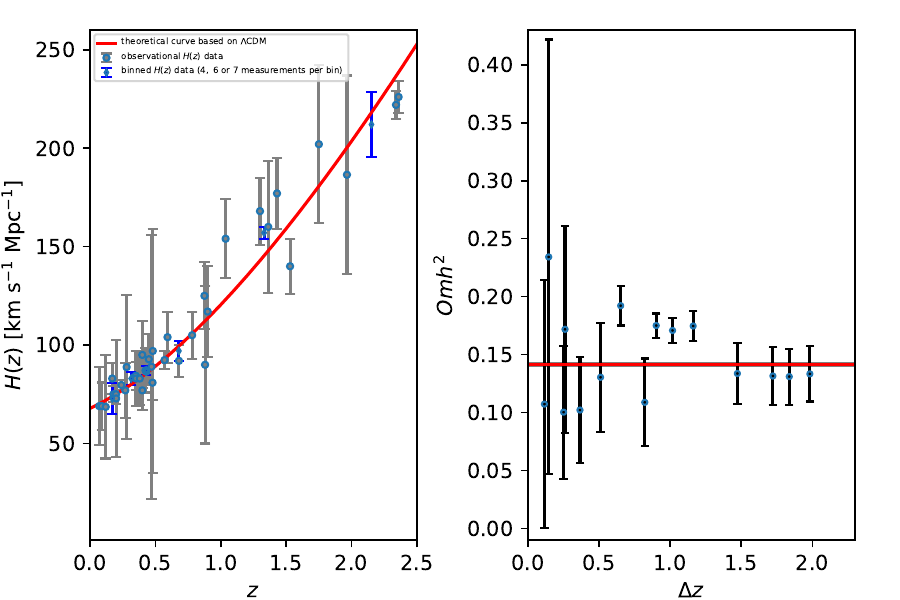}}\\
  \caption{37 reduced OHD associated with binned data and its corresponding $Omh^2$ diagnostic values for the median statistics technique. The panels (a), (b) (c) and (d) present the 3-4, 4-5, 4-5-6 and 4-6-7 measurements per bin case, respectively, where the red lines in the right panels represent the best-fit value of $Omh^2$ retrieved from the Planck CMB data.}\label{fig:HzOmh2}
\end{figure*}

\begin{figure*}[htbp]
\centering
  \subfloat[]{%
    \includegraphics[width=3in,height=1.6in]{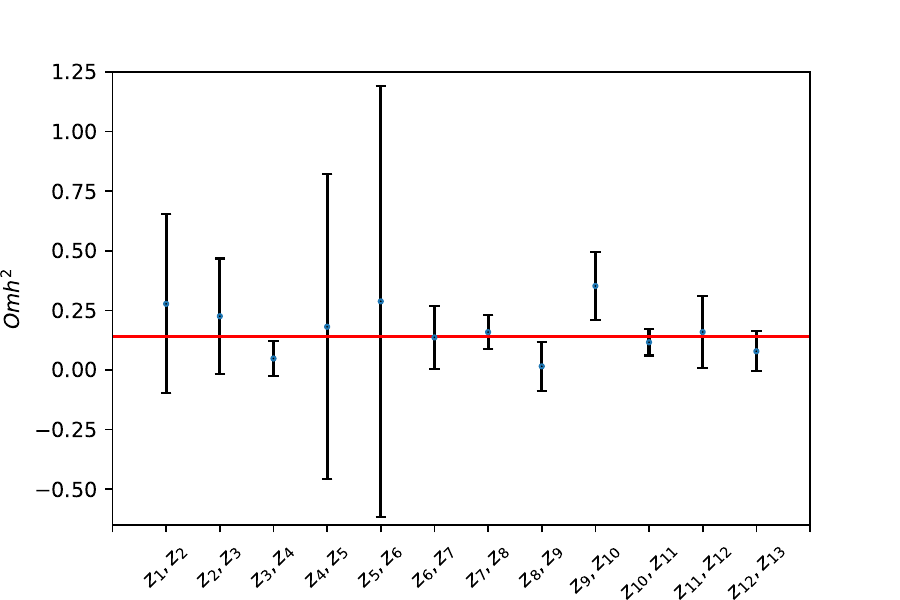}}
  \subfloat[]{%
    \includegraphics[width=3in,height=1.6in]{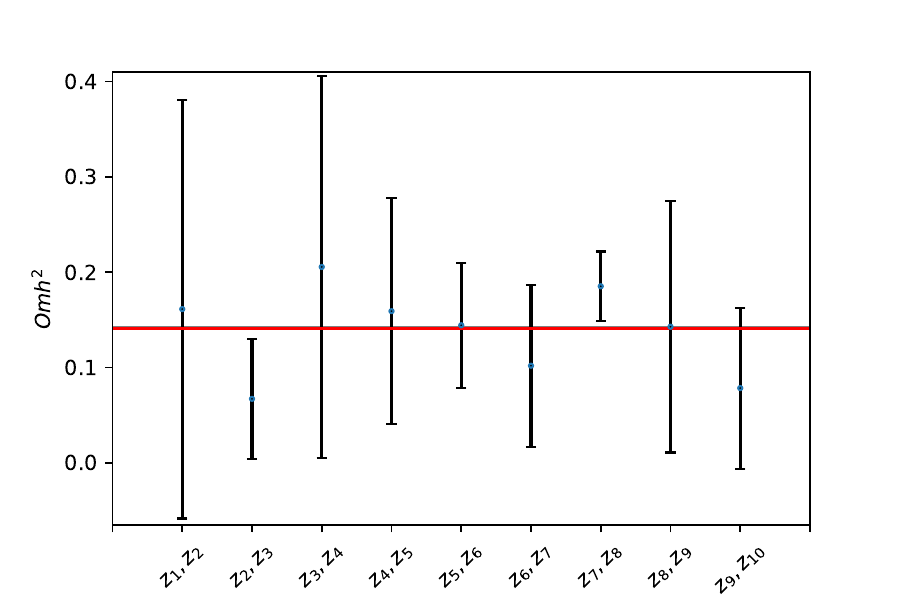}}\\
  \subfloat[]{%
    \includegraphics[width=3in,height=1.6in]{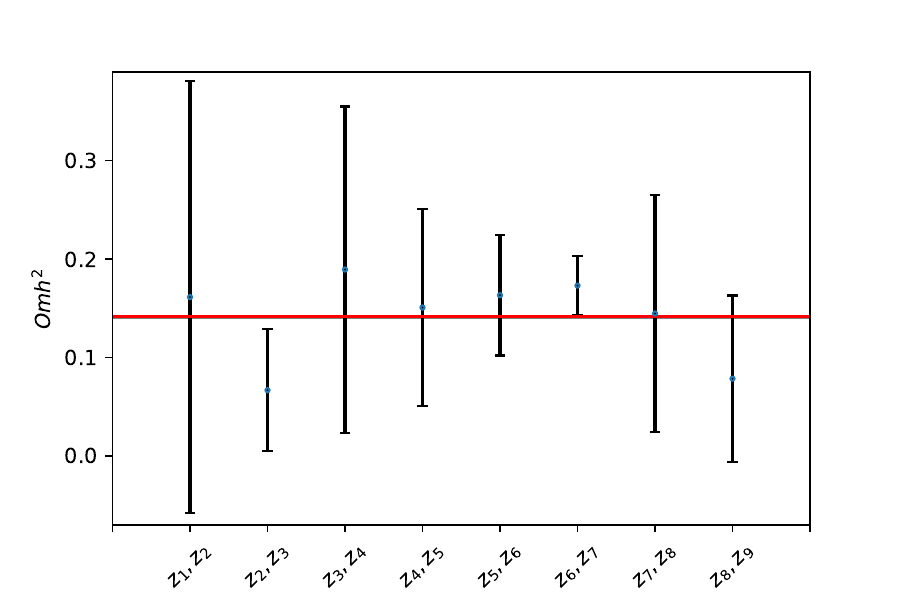}}
  \subfloat[]{%
    \includegraphics[width=3in,height=1.6in]{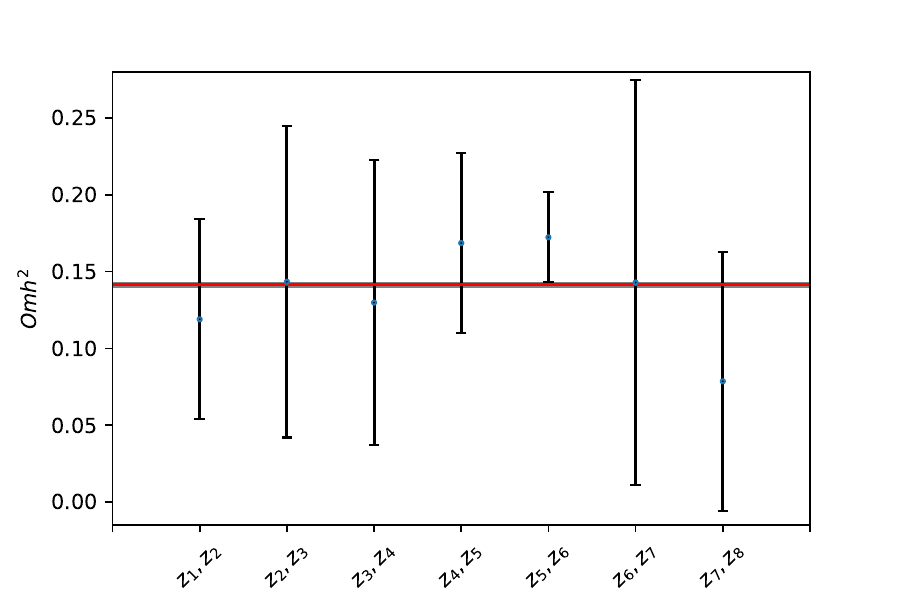}}\\
  \subfloat[]{%
    \includegraphics[width=3in,height=1.6in]{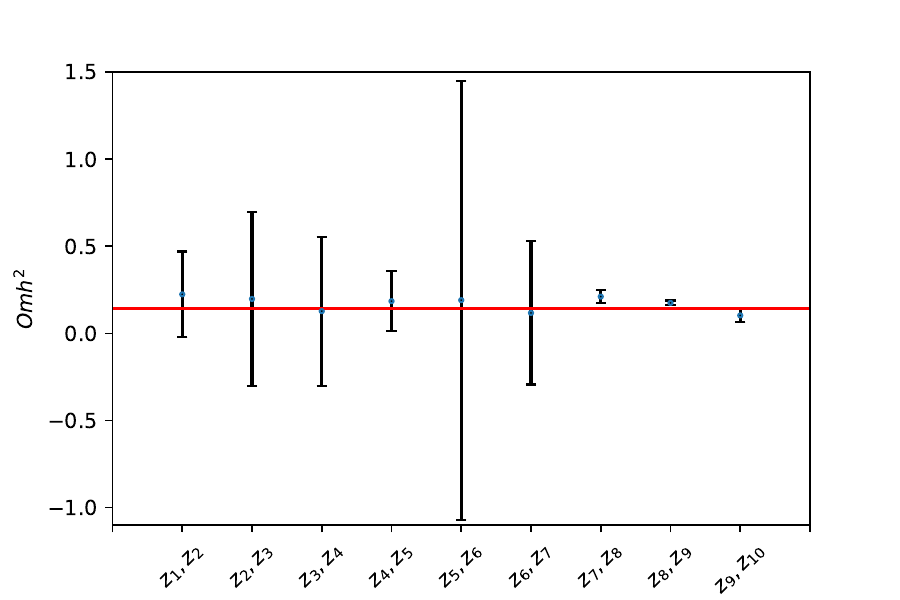}}
  \subfloat[]{%
    \includegraphics[width=3in,height=1.6in]{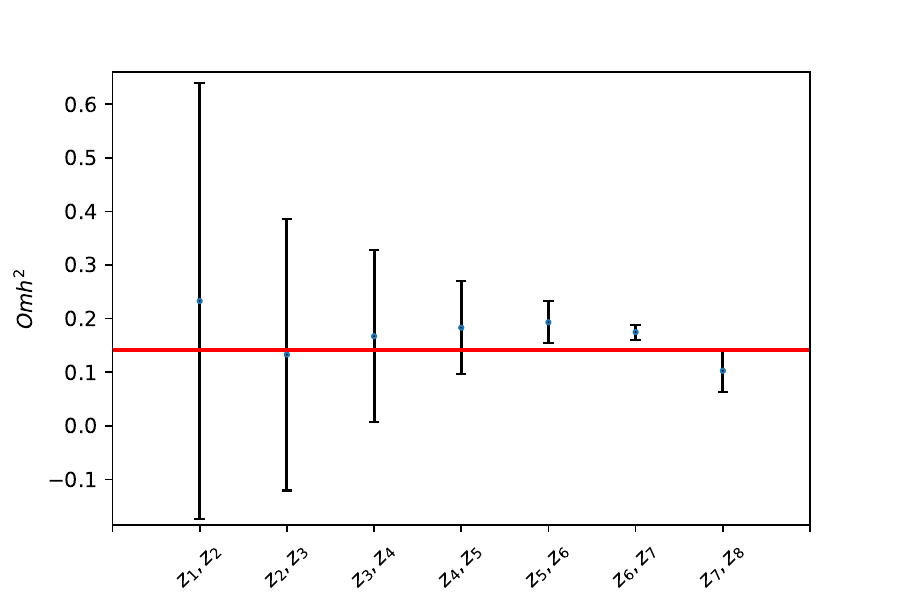}}\\
  \subfloat[]{%
    \includegraphics[width=3in,height=1.6in]{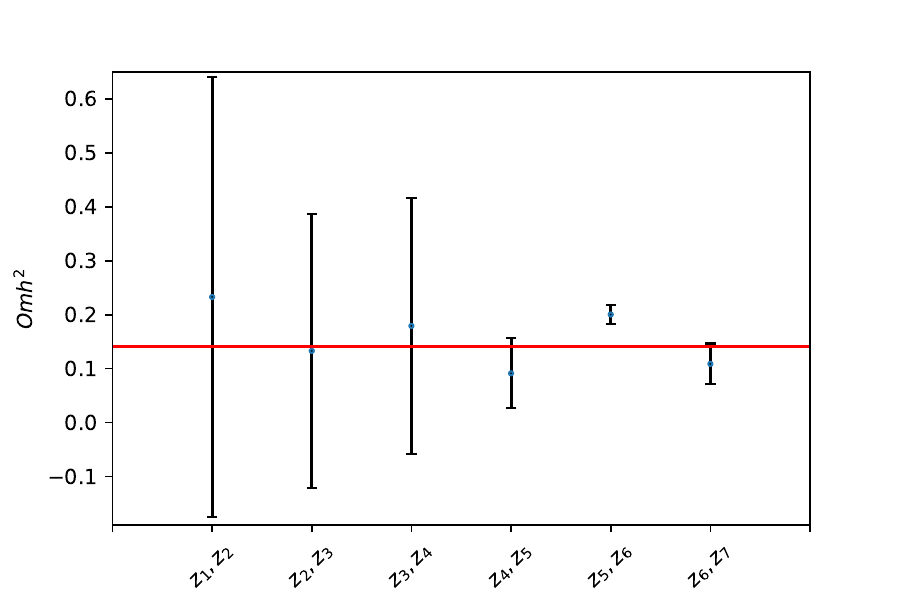}}
  \subfloat[]{%
    \includegraphics[width=3in,height=1.6in]{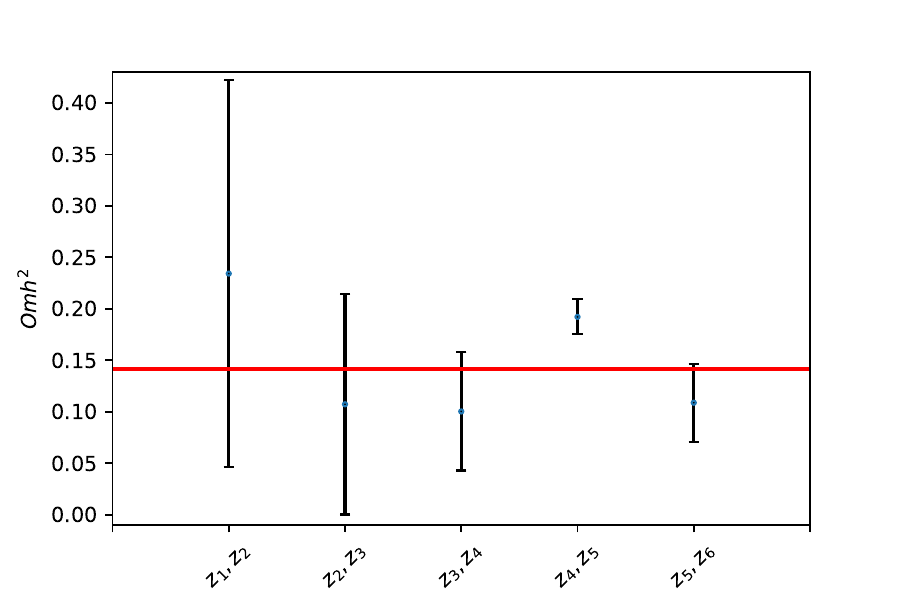}}\\
  \caption{$Omh^2$ diagnostic values for both the weighted mean and median statistics techniques with respect to continuous redshifts. The panels (a), (b), (c) and (d) present the weighted mean 3-4, 4-5, 5-6, and 5-6-7 measurements per bin, respectively. The panels (e), (f) (g) and (h) present the median statistics 3-4, 4-5, 4-5-6 and 4-6-7 measurements per bin case, respectively, where the red lines represent the best-fit value of $Omh^2$ retrieved from the Planck CMB data.}\label{fig:HzOmh3}
\end{figure*}

\begin{figure*}[htbp]
\centering
  \subfloat[]{%
    \includegraphics[width=6.8in,height=2.5in]{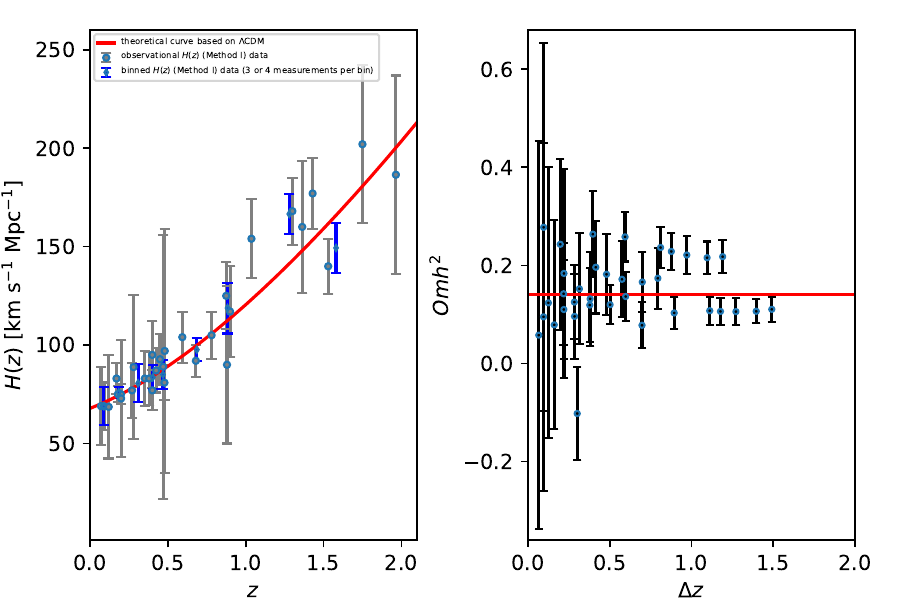}}\\
  \subfloat[]{%
    \includegraphics[width=6.8in,height=2.5in]{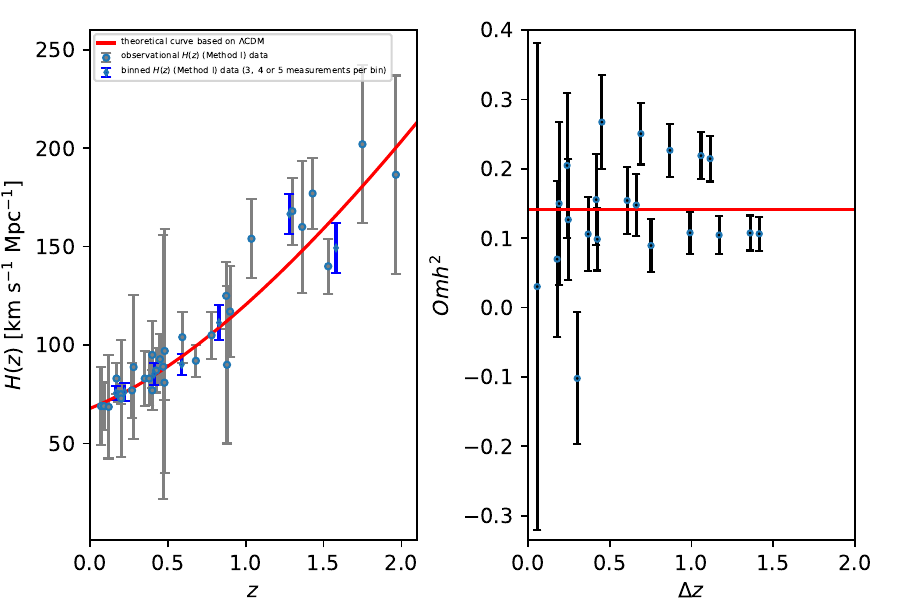}}\\
  \subfloat[]{%
    \includegraphics[width=6.8in,height=2.5in]{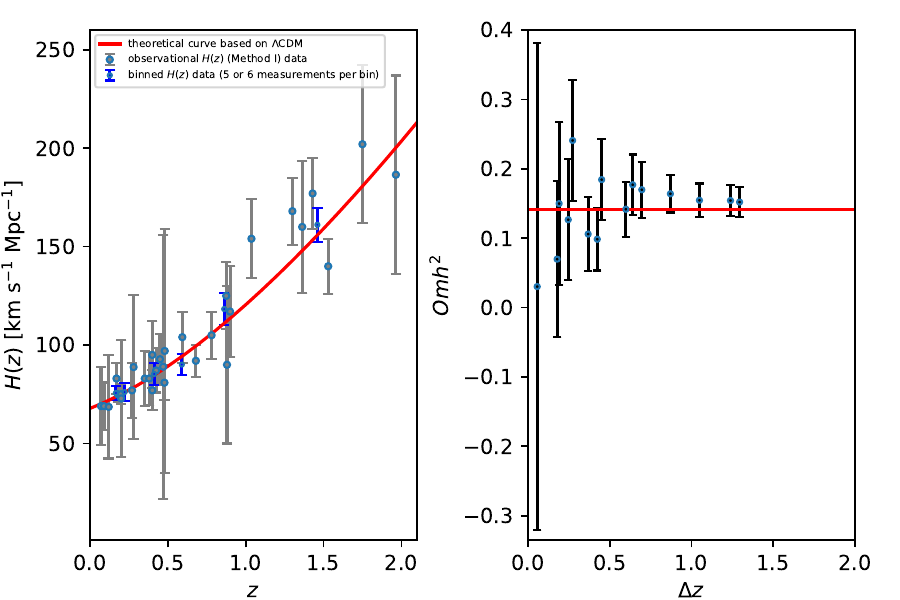}}\\
  \caption{OHD from the cosmic chronometer method associated with binned data and its corresponding $Omh^2$ diagnostic values for the weighted mean technique. The panels (a), (b) and (c) present the 3-4, 3-4-5 and 5-6 measurements per bin case, respectively, where the red lines in the right panels represent the best-fit value of $Omh^2$ retrieved from the Planck CMB data.}\label{fig:CCHzOmh1}
\end{figure*}

\begin{figure*}[htbp]
\centering
  \subfloat[]{%
    \includegraphics[width=6.8in,height=2.5in]{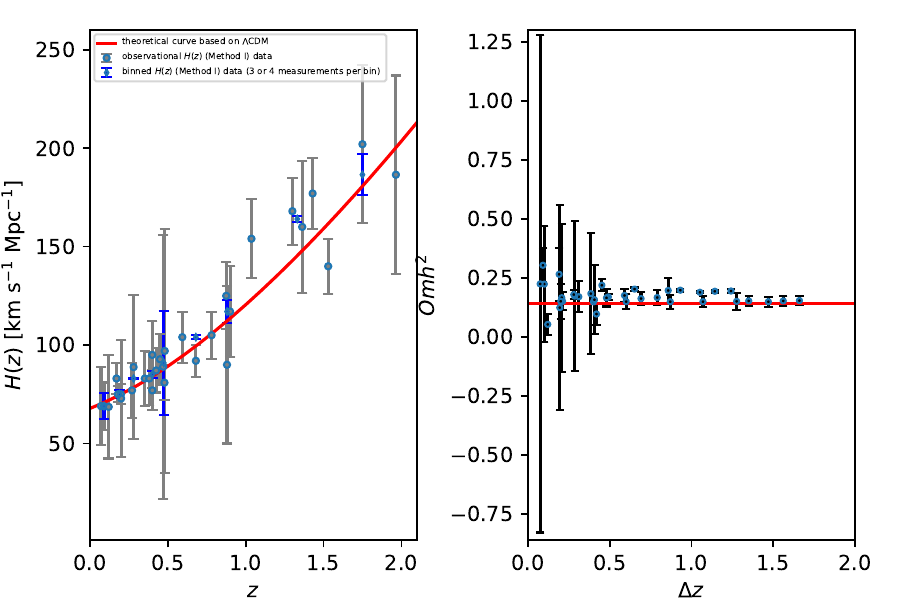}}\\
  \subfloat[]{%
    \includegraphics[width=6.8in,height=2.5in]{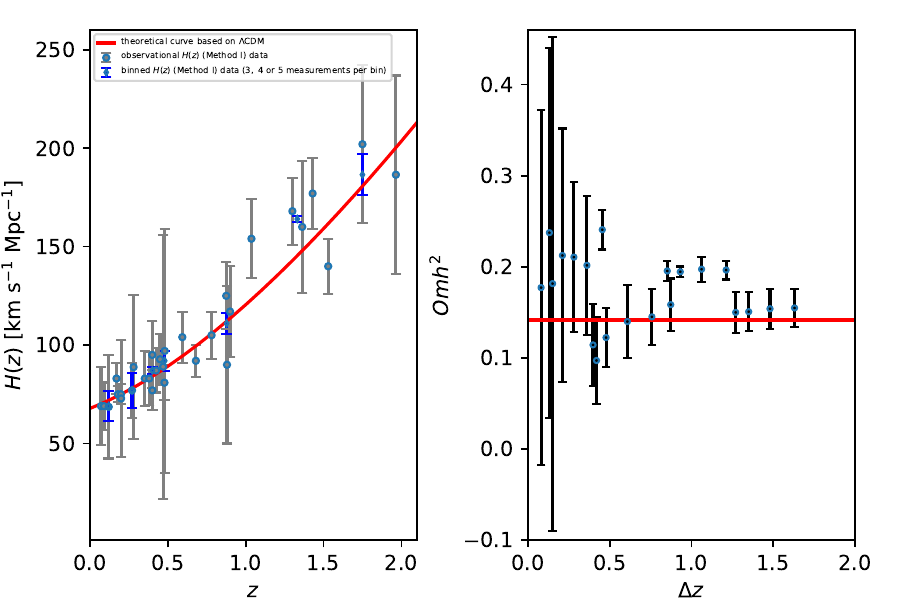}}\\
  \subfloat[]{%
    \includegraphics[width=6.8in,height=2.5in]{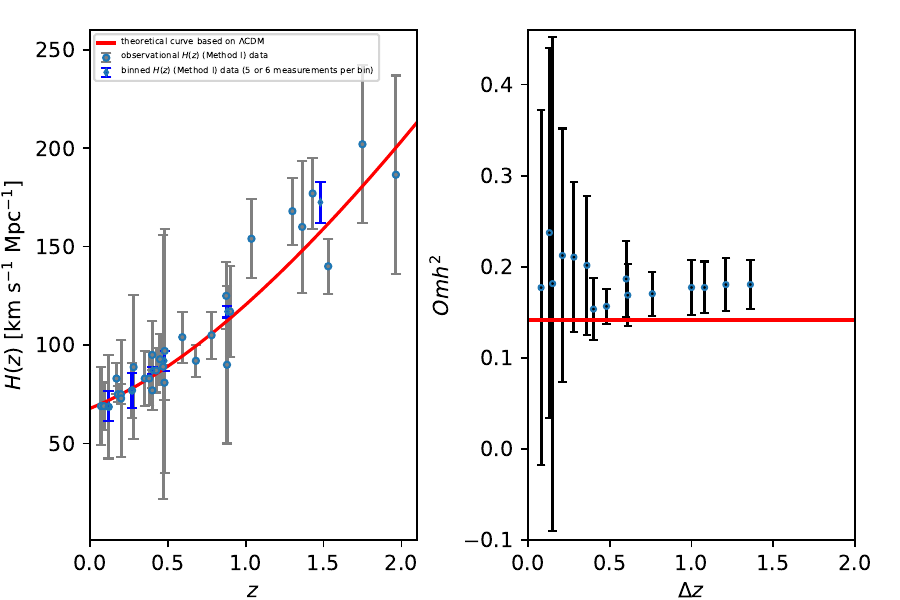}}\\
  \caption{Same as Fig. \ref{fig:CCHzOmh1}, but for the median statistics technique.}\label{fig:CCHzOmh2}
\end{figure*}

\begin{figure*}
\centering
  \subfloat[]{%
    \includegraphics[width=3.4in,height=2.5in]{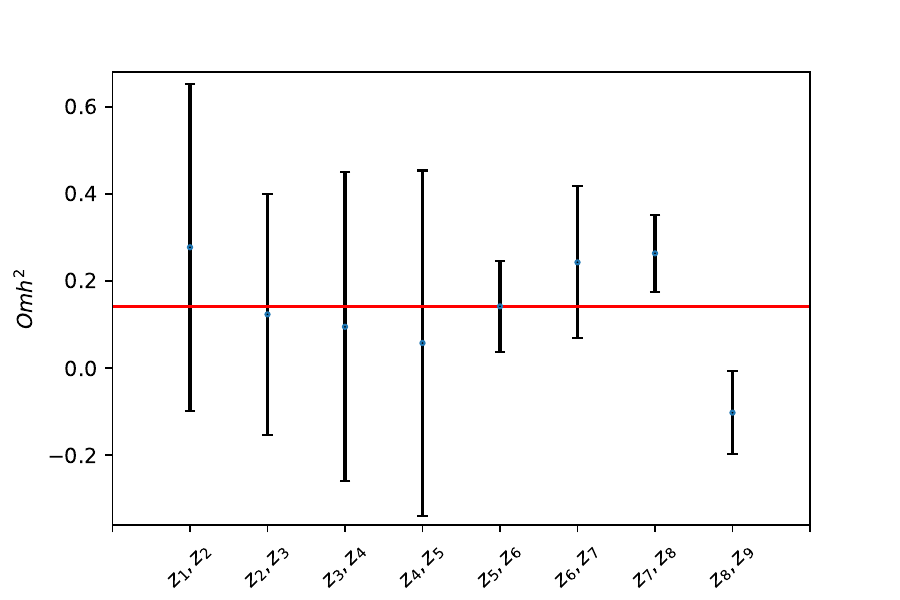}}
  \subfloat[]{%
    \includegraphics[width=3.4in,height=2.5in]{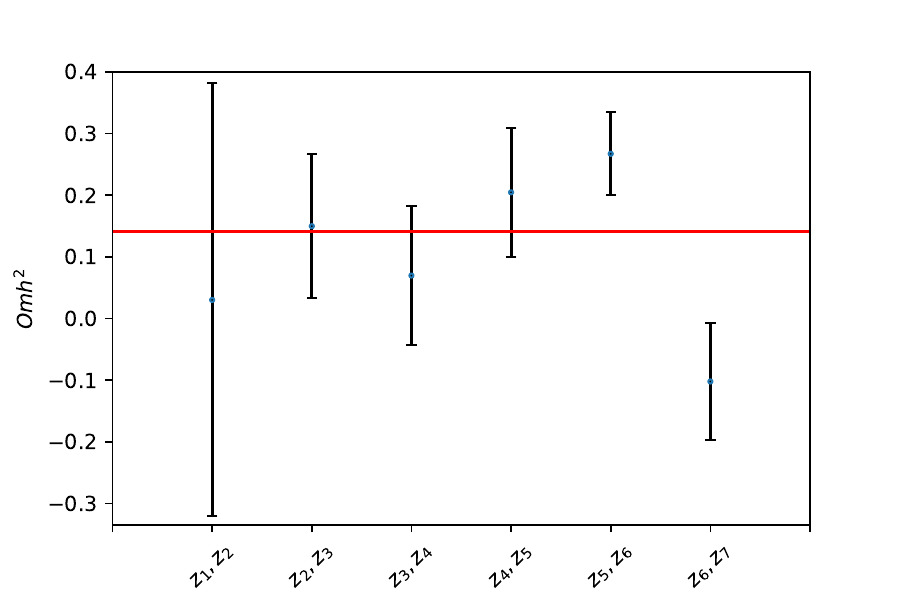}}\\
  \subfloat[]{%
    \includegraphics[width=3.4in,height=2.5in]{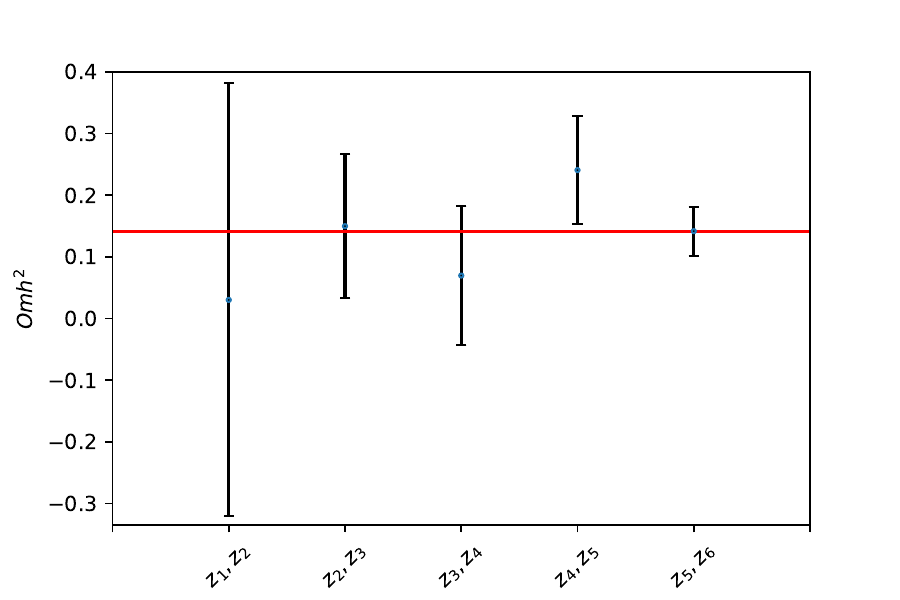}}
  \subfloat[]{%
    \includegraphics[width=3.4in,height=2.5in]{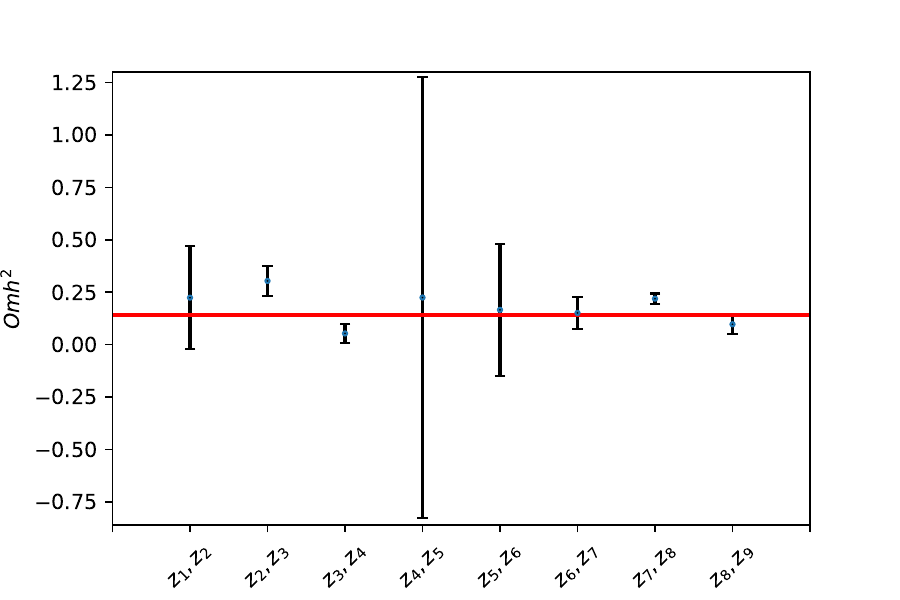}}\\
  \subfloat[]{%
    \includegraphics[width=3.4in,height=2.5in]{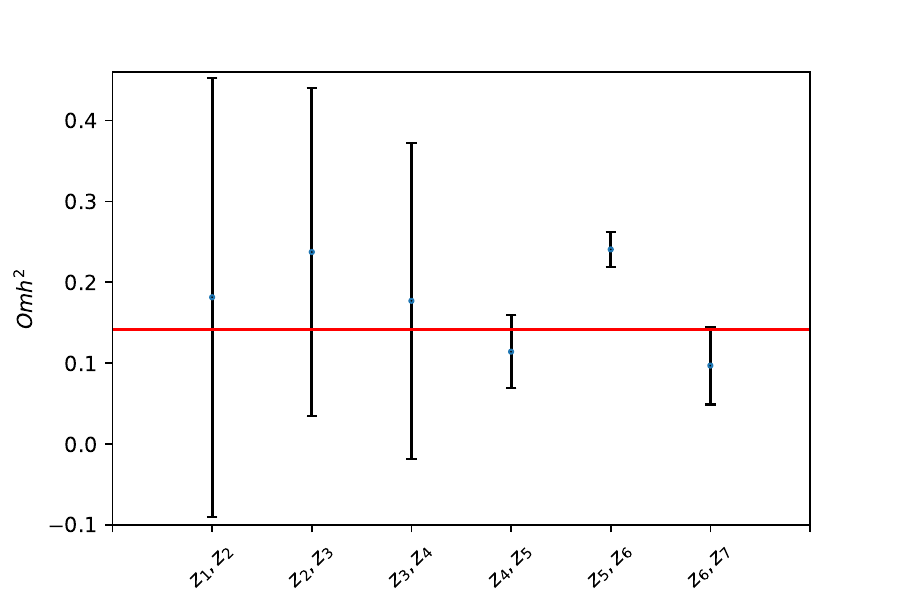}}
  \subfloat[]{%
    \includegraphics[width=3.4in,height=2.5in]{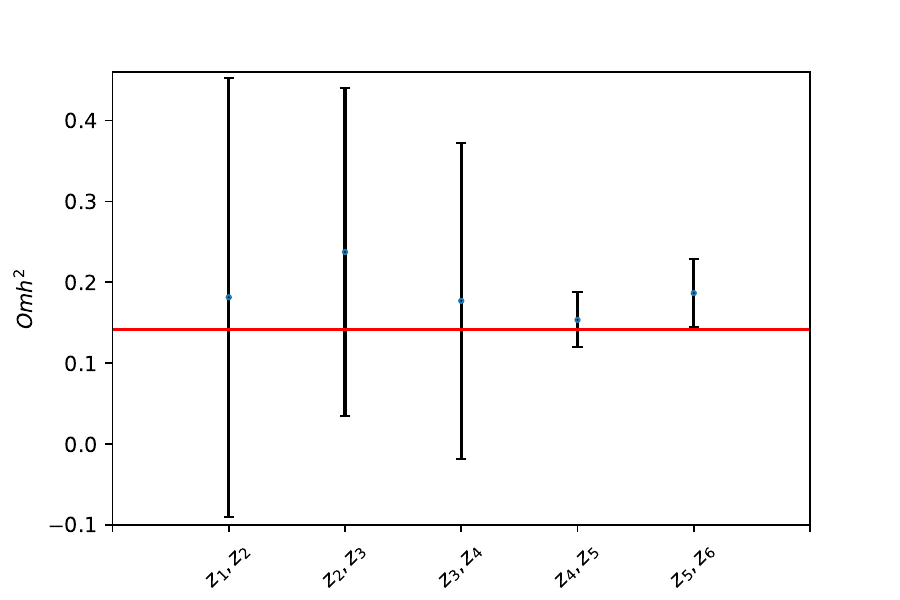}}\\
  \caption{Same as Fig. \ref{fig:HzOmh3}, but for OHD from the cosmic chronometer method. The panels (a), (b) and (c) present the weighted mean 3-4, 3-4-5, and 5-6 measurements per bin, respectively, and the panels (d), (e) and (f) present the median statistics 3-4, 3-4-5, 5-6 measurements per bin, respectively, where the red lines in the panels represent the best-fit value of $Omh^2$ retrieved from the Planck CMB data.}\label{fig:CCHzOmh3}
\end{figure*}

If $Om(z_{2};z_{1})$ is always a constant at any redshifts, then it demonstrates that the DE is of the cosmological constant
nature. In order to compare directly with the results from CMB, Sahni et al. \cite{2014ApJ...793L..40S} introduced a more convenient expression of the two-point
diagnostic, i.e.,
\begin{equation}
   \label{eq:Omh22points}
   Omh^2(z_{2};z_{1})=\frac{h^2(z_{2})-h^2(z_{1})}{(1+z_{2})^3-(1+z_{1})^3},
\end{equation}
where $h(z) = H(z)/100$ $\hunit$. The binned $H(z)$ points calculated based upon the aforementioned binning methods in each case therefore
yield $N(N-1)/2$ model-independent measurements of the $Omh^2(z_{2};z_{1})$ diagnostic, as shown in Figs. \ref{fig:HzOmh1}-\ref{fig:CCHzOmh3}, where the uncertainty $\sigma_{Omh^2(z_{2};z_{1})}$ can
be expressed as follows,
\begin{equation}\label{eq:deltaOmh2}
  \sigma^2_{Omh^2(z_{2};z_{1})}=\frac{4\left (h^2(z_2)\sigma^2_{h(z_2)}+h^2(z_1)\sigma^2_{h(z_1)}\right )}{\left ((1+z_2)^3-(1+z_1)^3\right )^2}.
\end{equation}

For $\Lambda$CDM, we have $Omh^2 \equiv \Om h^2$. The value of $\Om h^2$ is constrained tightly by the Planck observations to be
centered around 0.14 for the base of $\Lambda$CDM model fit \citep{2016AA...594A..13P}: the Planck temperature power spectrum data
alone gives $0.1426\\ \pm0.0020$, the Planck temperature data with lensing reconstruction gives $0.1415\pm0.0019$, and the Planck temperature
data with lensing and external data gives $0.1413\pm0.0011$, all at $1\sigma$ confidence level (CL). As stated in \cite{2016AA...594A..13P}, we
conservatively select $0.1415 \pm 0.0019$ as the Planck value.
If the $\Lambda$CDM model can hold, we would expect a constant value of $Omh^2$ at any redshift intervals. Sahni et al. \cite{2014ApJ...793L..40S} compared their results with the Planck value to check the validity of the $\Lambda$CDM model. From our perspective, we should not follow their goals to test the tension between our results and the Planck value. Instead, we should first consider whether the values of $Omh^2$ are constant or not.
Here we only use the Planck value for comparison purpose.

As shown in Fig. \ref{fig:HzOmh1}, the results from the weighted mean cases, within 1$\sigma$ confidence interval, are mostly continuous with both being constant (on average) and the Planck value, although some exceptions are presented and the best-fit $Omh^2(\Delta z)$ values fluctuate. Also, as shown in Fig. \ref{fig:HzOmh2}, the results from the median statistic cases are all continuous with both being constant and the Planck value within 1$\sigma$ confidence interval. Note that these results are not continuous redshift intervals, it only shows the differences for different $\Delta z$. Therefore, it would be useful if we plot the continuous results alone to extrapolate the outcomes. Then we illustrate these results with the binned OHD from both binning methods in Figs. \ref{fig:HzOmh3}, which shows that the values of $Omh^2$ for both binning methods fluctuate as the continuous redshift intervals change. The difference between the results from the weighted mean binning data and results from mean statistics is that the fluctuations from the former situation are more intense which makes the tendency more distinct from the first four panels of Fig. \ref{fig:HzOmh3}. Thus, it is fair to come up with the conclusion that the validity of $\Lambda$CDM is preserved.

Also, after binning the OHD from the cosmic chronometer method, the corresponding $Omh^2$ results with both binning techniques are demonstrated in Figs. \ref{fig:CCHzOmh1}-\ref{fig:CCHzOmh3}. It is evident that the fluctuations of the best-fit $Omh^2$ values are more intense than the results from full binned OHD for both binning methods. However, on average, the results are constant at 1$\sigma$ region. Hence, due to the two-point $Omh^2(z_2;z_1)$ diagnostic combined with binned OHD results, the $\Lambda$CDM model is favored. However, we note that the error bars of these results are much bigger than the Planck result, therefore we can only conclude that the flat $\Lambda$CDM model cannot be ruled out.

\section{Conclusions and Discussions} \label{sect:conclu}
In this paper, motivated by the investigations on the nature of DE, we test the validity of $\Lambda$CDM with the two-point $Omh^2(z_{2};z_{1})$ diagnostic by using 43 observational $H(z)$ data (OHD) which are obtained from the cosmic chronometers and BAO methods.

Firstly, instead of direct employment of the OHD on the $Omh^2(z_{2};z_{1})$ diagnostic, we introduce the two binning methods: the weighted mean and median statistics to reduce the noise in the data. After binning OHD, we conclude that the original OHD are not inconsistent with Gaussianity, and the binned data are all reasonable as Tables \ref{tab:wmbin} and \ref{tab:msbin} displayed. The OHD derived from the BAO method are not generally considered to be completely model-independent, even though as mentioned above the fiducial models indeed cannot affect the results (e.g., see Alam et al. \cite{2017MNRAS.470.2617A} P. 5), which means the data are model-independent after all. Nevertheless, due to the trust issue raised by some scientists, we also apply the binning method to the OHD derived from the cosmic chronometer method alone as listed in Tables \ref{tab:CCwmbin} and \ref{tab:CCmsbin} for comparison. The results all seem reasonable and, compared to the full binned OHD, the discrepancies are considerably small, which can be the indirect evidence for the validity of the OHD derived from the BAO method. Since the different measurements per bin do not significantly affect the results, we can acknowledge the robustness of the binning methods.

Secondly, combined with the set of binned OHD and, we exploit the $Omh^2(z_{2};z_{1})$ diagnostic to test if the $Omh^2$ values are constant. From Figs. \ref{fig:HzOmh1}-\ref{fig:CCHzOmh3}, we find that on average the $Omh^2$ values are mostly constant at 1 $\sigma$ confidence interval. Therefore, the flat $\Lambda$CDM model is not invalid. However, this does not mean that the dynamical DE models are not worth considering.

It is worth noticing that more independent OHD would bring more accuracy toward the binning methods, which can result in more reliable $Omh^2(z_{2};z_{1})$ values. Also, as the number of OHD grows, the binned OHD would be more efficient as a data clarification instrument that can be employed on cosmological constraints. OHD with higher precision and larger amounts are needed and valuable.

\begin{acknowledgements}
We thank the anonymous referee whose suggestions greatly helped us improve this paper. We thank Remudin Reshid Mekuria for helping us find the typo shown in the figures. This work was supported by National Key R\&D Program of China (2017YFA0402600),the National Natural Science Foundation of China (Grants No. 11573006, 11528306), the Fundamental Research Funds for the Central Universities and the Special Program for Applied Research on Super Computation of the NSFC-Guangdong Joint Fund (the second phase).
\end{acknowledgements}
\bibliographystyle{spphys}
\bibliography{Omh2}
\end{document}